\documentclass[12pt]{article}

\usepackage[labelfont=bf]{caption}
\usepackage{float}
\usepackage{graphicx}
\usepackage{bm}

\usepackage[super,numbers,sort&compress]{natbib}
\usepackage{amssymb}
\usepackage[hidelinks]{hyperref}
\usepackage{amsfonts}
\usepackage{color}
\usepackage{xcolor}
\usepackage{multicol}
\usepackage{xfrac} 
\usepackage{nicefrac} 
\usepackage{amsmath}  
\usepackage{comment}

\usepackage{mathrsfs}
\usepackage{resizegather}
\everymath{\displaystyle}

\emergencystretch=\maxdimen
\begin{document}
\baselineskip24pt

\title{}
{\noindent\large{\textbf {%
Resonance-Enhanced Four-Wave Mixing Imaging for Mapping Defect Regions in Vanadium-Doped WS$_2$ Monolayers
}}}
\\
\author{}
Felipe Menescal$^{1,\dagger}$, Frederico B. Sousa$^{1,2}$, Mingzu Liu$^{3,4}$, Ana P. M. Barboza$^{5}$, Igor F. Curvelo$^{5}$, Matheus J. S. Matos$^{5}$, Da Zhou$^{3}$, Bernardo R. A. Neves$^{1}$, Helio Chacham$^{1}$, Mauricio Terrones$^{3,4,6,7}$, Bruno R. Carvalho$^{8,\ast}$, Leandro M. Malard$^{1,\ast}$
\\
{\small%
$^{1}$Departamento de Física, Universidade Federal de Minas Gerais, Belo Horizonte, Minas Gerais 30123-970, Brazil\\
$^{2}$Departamento de F\'isica, Universidade Federal de São Carlos, São Carlos, São Paulo 13565-905, Brazil\\
$^{3}$Department of Physics, The Pennsylvania State University, University Park, PA 16802, United States of America\\
$^{4}$Center for 2-Dimensional and Layered Materials, The Pennsylvania State University, University Park, PA 16802, United States of America\\
$^{5}$Departamento de Física, Universidade Federal de Ouro Preto, Ouro Preto, Minas Gerais, 35400-000, Brazil\\
$^{6}$Department of Chemistry, The Pennsylvania State University, University Park, PA, 16802, United States of America\\
$^{7}$Department of Materials Science and Engineering, The Pennsylvania State University, University Park, Pennsylvania, 16802, United States of America\\
$^{5}$Departamento de Física Teórica e Experimental, Universidade Federal do Rio Grande do Norte, Natal, Rio Grande do Norte 59078-970, Brazil\\
$^{\dagger}$Present address: Laboratoire de Physique de l’Ecole normale supérieure, ENS, Université PSL, CNRS, Sorbonne Université, Université de Paris, 24 rue Lhomond, 75005 Paris, France\\
$^{\ast}$Corresponding authors: brunorc@fisica.ufrn.br, lmalard@fisica.ufmg.br\\
}

\section*{Abstract}

\noindent \textbf{%
Defect engineering is crucial for tuning 2D transition metal dichalcogenide properties for quantum and optoelectronic applications. While conventional photoluminescence (PL) and Raman spectroscopies are important characterization tools, their mapping in large area samples can be time-consuming and lacks direct sensitivity for comprehensive defect characterization. Here, we introduce resonance-enhanced four-wave mixing (FWM) imaging for precise imaging and characterization of vanadium-induced defect states in WS$_2$ monolayers. Our multi-modal investigation, integrating hyperspectral PL, Raman, and supported by density functional calculations, reveals nanoscale doping inhomogeneities, their influence on excitonic and vibrational properties.  We observe resonance-enhanced FWM signals correlating with vanadium-induced defect regions, evidencing their unique nonlinear optical response. This work establishes FWM as an essential platform for high-resolution, defect-sensitive imaging, advancing defect-engineered excitonic devices and enabling novel nonlinear quantum photonics.
}

\section*{Introduction}
Two-dimensional (2D) materials hold immense promise for next-generation quantum technologies, yet the precise spatial control and characterization of defects, which critically influence their electronic and optical behavior, remain central challenges~\cite{lin2016defect,liang2021defect,ippolito2022defect,OrtizJimenez2023,Lee2023}.
In transition metal dichalcogenides (TMDs), the ability to engineer and probe defect states is critical for developing quantum materials, as defects modify band structures, excitonic interactions, and spin-valley coupling~\cite{Zhang2018,Cai2020,Nguyen2021,Nisi2022,Zhang2023,sousa2024effects,sousa2024giant,sousa2025strong}. Specifically, vanadium doping in WS$_2$ introduces localized electronic and magnetic states~\cite{zhang-VWS2,Zhang2023,sousa2024effects}, enabling novel excitonic phenomena such as defect-assisted transitions and altered valley polarization, highly relevant for spintronic and valleytronic applications.

While photoluminescence and Raman spectroscopy have been widely employed to study defect-related spectral features~\cite{review-apr}, hyperspectral mapping using these linear techniques can be time-consuming and often lack the direct sensitivity for a comprehensive understanding of defect states. These limitations require advanced nonlinear optical techniques capable of probing localized interactions with high spatial resolution and sensitivity, offer fast acquisition times and are suitable for large area samples~\cite{review-zhipei, science,carvalho-dark,sousa2025nonlinear}. It is important to note that most nonlinear optical imaging methods developed relies on second-order process, but few works have used FWM \cite{patrick, lafeta-nonlinear} but with focus on exciton resonances instead of defect characterization.

Here we show that resonance-enhanced four-wave mixing imaging provides unprecedented insights into the unique nonlinear optical responses of vanadium-induced defect states in WS$_2$ monolayers. Unlike linear spectroscopies, FWM sensitively reveals subtle changes in nonlinear optical responses associated with defects states. By integrating FWM imaging with hyperspectral PL and Raman mappings, and complementing these results with density functional theory calculations, we construct a comprehensive framework for understanding the nanoscale interplay between doping, strain, and exciton states. Our findings position this nonlinear optical imaging approach, particularly FWM, as an essential technique for advanced defect characterization in 2D semiconductors, significantly enabling the development of next-generation photonic and quantum technologies.

\section*{Results and Discussion} 

\subsection*{Photoluminescence mapping of doping inhomogeneity}
\begin{figure}[!htbp]
    \centering
    \includegraphics[width=0.85\textwidth]{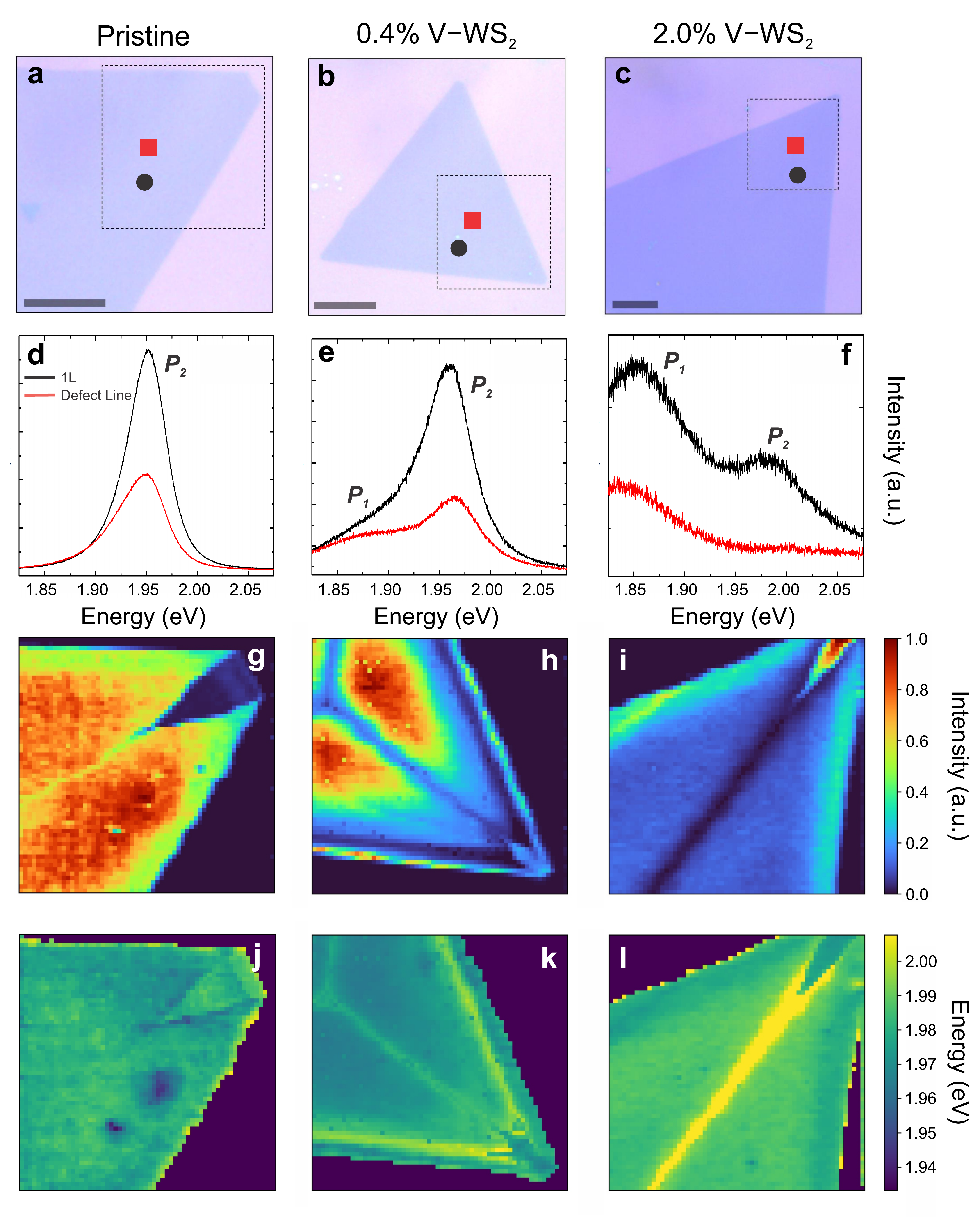}
    \caption{{\bf Optical imaging and PL mapping of pristine and vanadium-doped WS$_2$ monolayers.} Optical images of {\bf a.} pristine, {\bf b.} 0.4~at\%, and {\bf c.} 2.0~at\% V-doped WS$_2$ monolayers. Scale bars represent 10~$\mu$m. 
    Single PL spectra acquired from monolayer (1L, black curve) and defect line (red curve) regions for {\bf d.} pristine, {\bf e.} 0.4~at\%, and {\bf f.} 2.0~at\% V-doped samples. 
    PL intensity maps (integrated area) of $P_2$ peak for: {\bf g.} pristine, {\bf h.} 0.4~at\%, and {\bf i.} 2.0~at\% V-doped. PL energy maps of $P_2$ peak for: {\bf j.} pristine, {\bf k.} 0.4~at\%, and {\bf l.} 2.0~at\% V-doped. Map area: 20$\times$20 $\mu$m$^2$.
    }
    \label{fig1}
\end{figure}

%
Pristine and vanadium (V)-doped WS$_2$ monolayers were synthesized using a single-step chemical vapor deposition (CVD) process, following methods reported by Zhang~\textit{et al.}~\cite{zhang-VWS2} (see Methods). The pristine sample contains 0 (zero) atomic percent (at\%) of vanadium (Figure~\ref{fig1}a), while the doped samples exhibit V concentrations of approximately 0.4~at\% and 2.0~at\% (Figure~\ref{fig1}b,c). This study primarily focuses on the 2.0~at\% doping level, a critical regime where the ferromagnetic response is maximized~\cite{zhang-VWS2} and defect-induced optical phenomena are most pronounced~\cite{sousa2024giant,review-apr}. The atomic vanadium concentrations in each sample were consistent with previous characterizations reported elsewhere~\cite{zhang-VWS2}.

Figure~\ref{fig1}d-f shows the PL spectra acquired from two distinct regions: a uniform monolayer domain (black circle) and a prominent defect line (red square), as visually marked in the optical images (Figure~\ref{fig1}a-c). In the pristine monolayer, a single emission peak is observed at approximately 1.97~eV, characteristic of the A exciton (Figure~\ref{fig1}d). Upon vanadium doping, this A exciton peak undergoes a notable intensity reduction, significant linewidth broadening, and energy blueshift, evolving into the $P_2$ peak. Meanwhile a new low-energy peak, designated $P_1$, emerges only in the doped samples (Figure~\ref{fig1}e,f). These two peaks are attributed to radiative recombination from a vanadium-induced donor-like state ($P_1$) and transitions from the conduction band minima ($P_2$)~\cite{sousa2024effects}, respectively. A comprehensive discussion of their origin is detailed in prior work~\cite{sousa2024effects}. 

To spatially resolve the influence of vanadium doping on emission uniformity, we performed a hyperspectral PL mapping across 20$\times$20~$\mu$m$^2$ region, corresponding to the dashed-square areas in Figure~\ref{fig1}a-c. 
The A exciton and $P_2$ intensity maps (Figure~\ref{fig1}g-i) strikingly reveal a pronounced decrease in emission along the characteristic dark line extending from the monolayer center to its vertex. This observation is consistent with enhanced nonradiative recombination at these structural defect sites, which act as efficient exciton quenchers~\cite{rosa-WS2}.
Complementary energy maps show that while the A exciton in the pristine sample exhibits spatial uniformity (Figure~\ref{fig1}j), the $P_2$ emission in doped samples displays a systematic blueshift along the defect line (Figure~\ref{fig1}k,l). In contrast, the $P_1$ peak, indicative of vanadium-induced states, exhibits a clear redshift and darker contrast along this same defect line (see Figure~S1), strongly suggesting preferential incorporation of vanadium dopants at these growth-induced defect regions during synthesis~\cite{rosa-WS2,sousa2024effects}. These PL findings highlight the significant spatial inhomogeneity introduced by vanadium doping.

Further physical characterization using scanning probe microscopy (SPM) (see Methods and Figure~S2 for details) confirmed that these defect lines exhibit distinct structural, electronic, and mechanical properties. These SPM findings strongly corroborate the preferential incorporation of vanadium dopants along these growth-induced lines, directly correlating with the observed localized PL features.

\subsection*{Raman mapping and strain effects}
\begin{figure}[!htpb]
    \centering
    \includegraphics[width=1.0\textwidth]{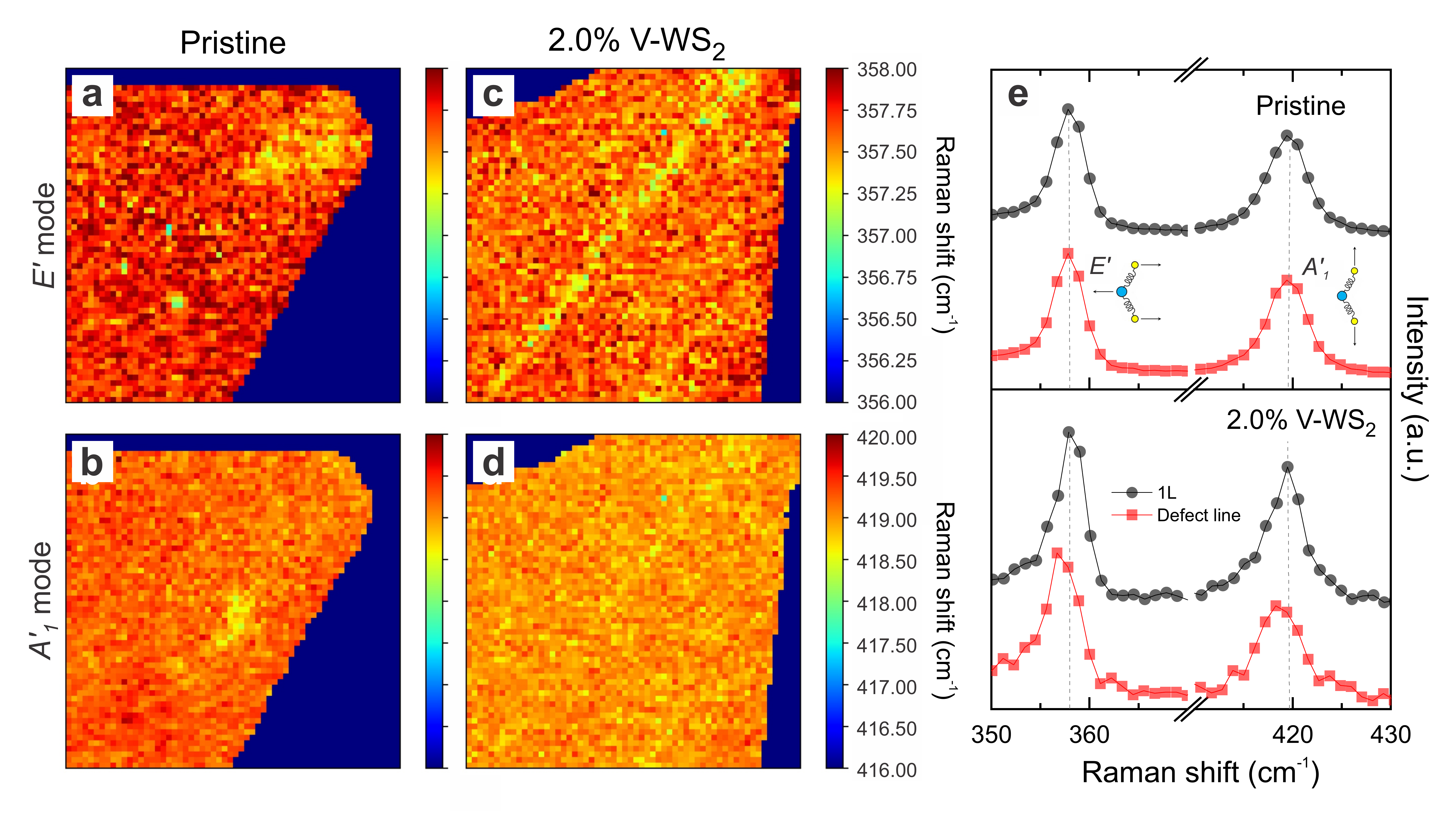}
    \caption{{\bf Raman hyperspectral maps of pristine and vanadium-doped WS$_2$ monolayers.}
    {\bf a,c.} $E^{\prime}$ and {\bf b,d.} $A^{\prime}_{1}$ frequency maps for pristine and 2.0~at\% V-doped WS$_2$, respectively. All maps were acquired using a 2.71~eV excitation over a 20$\times$20~$\mu$m$^2$ area. 
    {\bf e.} Raman spectra comparing monolayer (1L, black circles) and the defect line (red squares) regions for pristine and 2.0~at\% V-doped samples. All Raman data were normalized to the Si peak at 521.6~cm$^{-1}$.}
    \label{fig2}
\end{figure}

%
Hyperspectral Raman measurements were performed on pristine, 0.4~at\%, and 2.0~at\% V-doped WS$_2$ monolayers using a 2.71~eV excitation. We focused the analysis on the same regions previously investigated with PL mapping (Figure~\ref{fig1}). For pristine sample, the hyperspectral frequency maps of the first-order $E^{\prime}$ and $A^{\prime}_{1}$ modes (Figure~\ref{fig2}a,b) reveal negligible variations in both intensity and frequency, indicating highly uniform vibrational properties. A similar homogeneous Raman response was observed for the 0.4~at\% V-doped WS$_2$ monolayer (see Figure~S3 and Figure~S4 for intensity maps). 

In stark contrast, the 2.0~at\% V-doped sample exhibits distinct spatial variations specifically along the defect line (Figure~\ref{fig2}c,d). Here, the $E^{\prime}$ mode displays a more pronounced redshift compared to the $A^{\prime}_{1}$ mode. This behavior is mainly attributed to strain effects induced by vanadium doping, which can alter the effective mass and reduce the metal-sulfur bond length within the WS$_2$ lattice~\cite{WS2-raman-modes,dadgar2018strain}. Quantitatively, the observed shifts of approximately 1~cm$^{-1}$ for $E^{\prime}$ and 0.6--0.8~cm$^{-1}$ for $A^{\prime}_{1}$ (Figure~\ref{fig2}e) are consistent with local strain levels ranging from 1.0--1.5\%~\cite{dadgar2018strain,dos2023intrinsic}. Crucially, these Raman results strongly corroborate the PL findings, thereby confirming that the identified defect line corresponds to regions with enhanced vanadium doping and associated structural strain.

\subsection*{Four-wave mixing imaging of defect states}
%
While conventional PL and Raman spectroscopies provide valuable insights into defect distribution and local strain, hyperspectral mapping with these linear techniques can be time-consuming, and they often lack the direct sensitivity to the nonlinear optical responses and coherent excitonic interactions crucial for a complete understanding of defect states. Nonlinear optical techniques, and four-wave mixing (FWM) imaging in particular, offer a distinct advantage by providing enhanced sensitivity to these subtle, localized interactions and enabling faster acquisition times~\cite{carvalho-dark,sousa2025nonlinear,lafeta-nonlinear,sousa2024effects}. In addition, resonance-enhanced FWM imaging, as applied in this work, provides unprecedented insights into their unique nonlinear optical responses. This technique selectively probes defect-induced excitonic states by enhancing specific transitions sensitive to the local nonlinear susceptibility ($\chi^{(3)}$), thereby enabling spectroscopic access to unique defect-signatures.

\begin{figure}[!htbp]
    \centering
    \includegraphics[width=0.8\textwidth]{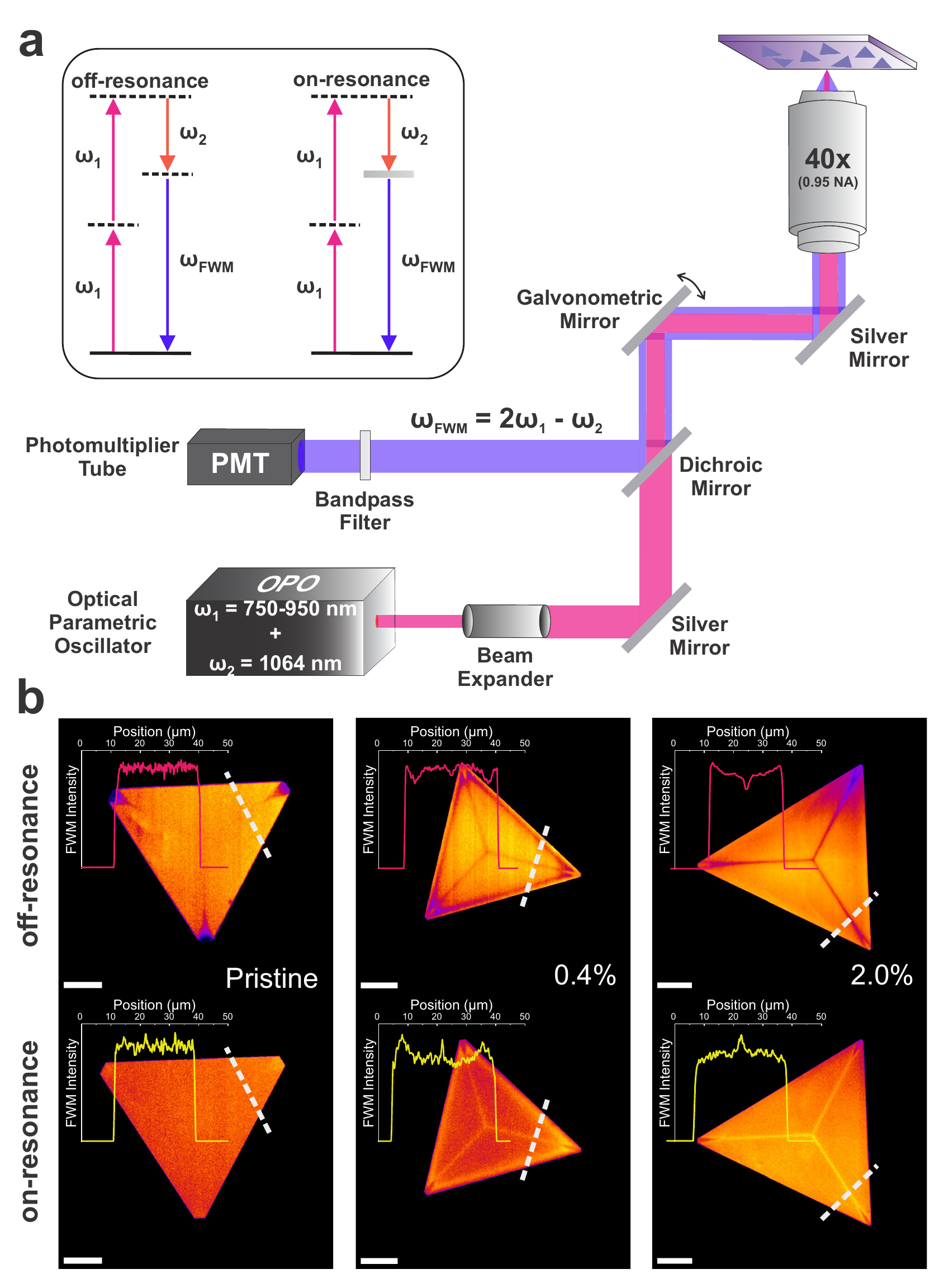}
    \caption{{\bf FWM imaging of pristine and vanadium-doped WS$_2$ monolayers.}
    {\bf a.} Schematic of the FWM experimental setup. Further details are in the Methods section. The inset shows energy diagrams of the FWM generation process, illustrating off-resonance (left) and on-resonance (right) conditions.
    {\bf b.} FWM images of pristine, 0.4~at\%, and 2.0~at\% V-doped WS$_2$ monolayers with FWM signal at 2.03~eV (off-resonance, top row) and 1.82~eV (on-resonance, bottom row), respectively. Scale bars are 20~$\mu$m.}
    \label{fig4}
\end{figure}

In our degenerate FWM experiment (Figure~\ref{fig4}a), the interaction of two incident photons of tunable frequency ($\omega_{1}$) and a third photon at a fixed wavelength (1064~nm, $\omega_{2}$) generates a signal at $\omega_{\text{FWM}} = 2\omega_{1} - \omega_{2}$~\cite{boyd2008nonlinear,lafeta-nonlinear,lafeta2017anomalous,hendry2010coherent}.
Our experimental setup permits tuning of $\omega_{\text{FWM}}$ from 544 to 858~nm, thereby covering the critical WS$_2$ A exciton resonance (610--680~nm). By systematically scanning $\omega_{1}$ across excitonic energies, we observe significantly enhanced FWM signals when $\omega_{\text{FWM}}$ matches real excitonic states, providing a resonance-enhanced optical contrast (Figure~\ref{fig4}a, inset). 

Figure~\ref{fig4}b shows real-time FWM images of pristine, 0.4~at\%, and 2.0~at\% V-doped WS$_2$ samples at two distinct FWM signal energies: 2.03~eV (610~nm, off-resonance) and 1.82~eV (680~nm, on-resonance with the $P_1$ defect-related peak).

Notably, in the doped samples, the FWM response is strongly modulated along the previously identified defective lines by resonant excitation. At the off-resonant energy (2.03~eV, top row), both 0.4~at\% and 2.0~at\% V-doped samples exhibit a distinct FWM quenching along the defective lines, manifesting as a dark contrast. The 0.4~at\% sample additionally shows dark contrast near flake edges. Conversely, when the FWM signal is tuned to the resonant energy of 1.82~eV (bottom row), we observe a striking enhancement of FWM emission along these same regions, appearing with bright contrast. This pronounced behavior directly correlates with the $P_1$ PL peak, which is associated with vanadium-induced defect states~\cite{rosa-WS2,sousa2024effects}. 

The FWM images clearly delineate two spatially distinct regions: (i) the one-dimensional bisection lines (from center to vertices), which our prior PL and SPM results confirmed as preferential sites for dopant incorporation, and (ii) the surrounding monolayer flake area. The observed reduction in FWM intensity along defect lines at 2.03~eV, which contrasts sharply with the stronger FWM signals in these same defect regions at 1.82~eV of V-doped samples, provides compelling direct evidence of increased vanadium-induced defect states. These observations demonstrate that vanadium-induced defect states strongly modulate nonlinear excitonic interactions, most likely by significantly enhancing third-order nonlinear susceptibility $\chi^{(3)}$, localized at these defect sites. 

To quantify this contrast difference between on- and off-resonance FWM images, we extracted line profiles (indicated by dashed white line in insets of Figure~\ref{fig4}b) across the defect lines and adjacent monolayer regions. The FWM intensity was normalized to the monolayer signal. We define a defect line (DL) contrast factor (C) as: $C=\frac{I_\text{DL}-I_\text{1L}}{I_\text{DL}+I_\text{1L}}$, where $I_\text{DL}$ and $I_\text{1L}$ are the FWM intensities from the defect line and monolayer regions, respectively. Analysis of these line profiles reveals an approximately 3-fold contrast enhancement for the 0.4~at\% V-doped sample, $\left| \frac{C_\text{on}}{C_\text{off}}\right| = 3.4 \pm 0.5$. In contrast, the 2.0~at\% doped sample exhibits a more modest enhancement of about 1-fold, $\left| \frac{C_\text{on}}{C_\text{off}}\right| = 1.3 \pm 0.3$, likely due to saturation effects in the off-resonance signal at higher doping concentrations.

\begin{figure}[!htbp]
    \centering
    \includegraphics[width=\textwidth]{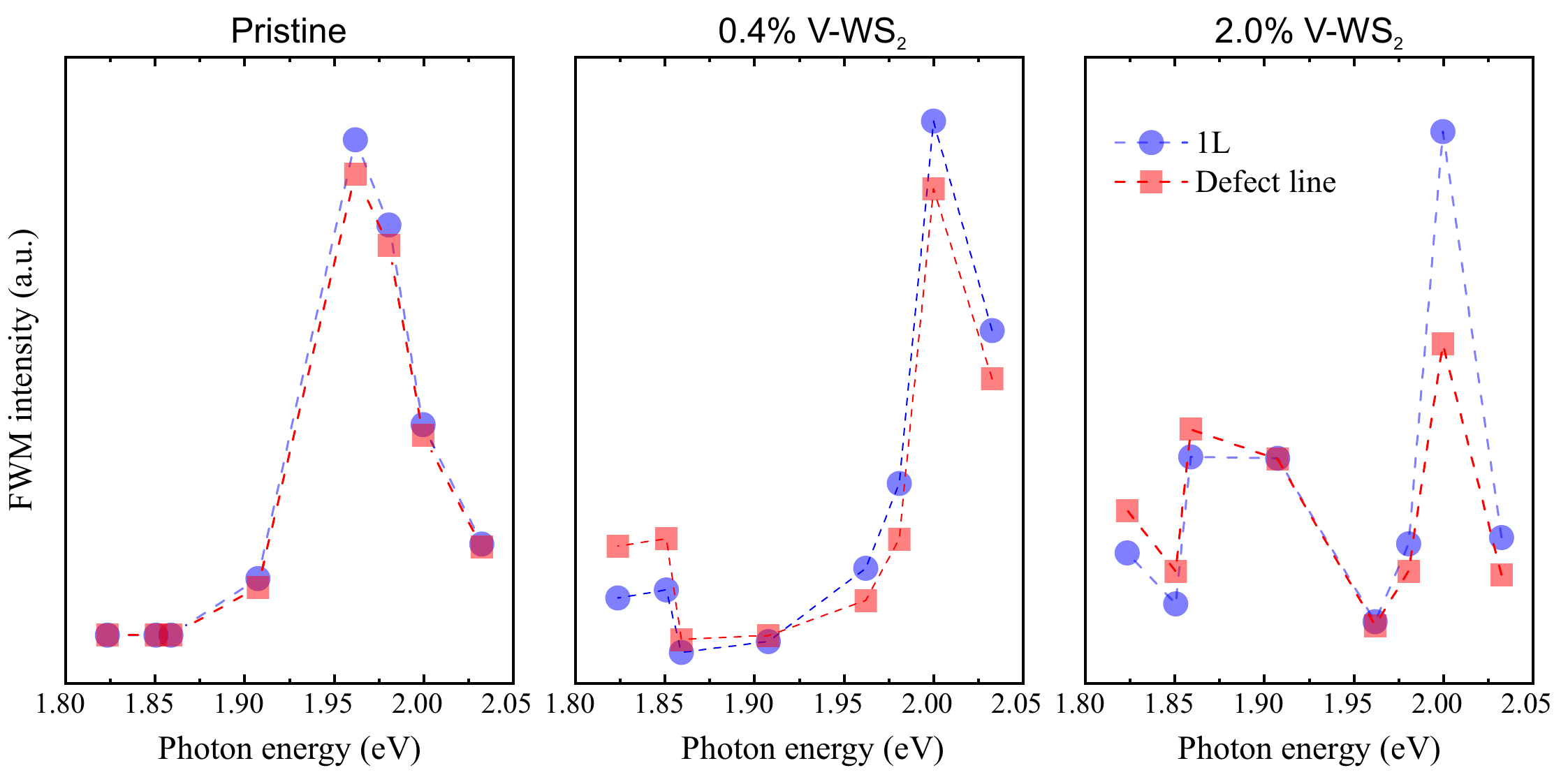}
    \caption{{\bf Resonance-enhanced FWM response of pristine and vanadium-doped WS$_2$ monolayers}. FWM excitation profiles for pristine, 0.4~at\%, and 2.0~at\% doped WS$_2$ samples acquired from the monolayer (red squares) and defect line (blue circles) regions. Intensities are normalized by the quartz reference signal at each excitation energy.}
    \label{fig5}
\end{figure}

To further investigate the spectral characteristics of these FWM resonances, Figure~\ref{fig5} shows the FWM intensity as a function of the generated photon energy (ranging from 1.82--2.03~eV) for pristine, 0.4~at\%, and 2.0~at\% doped WS$_2$ samples. Data was acquired from both the monolayer (red squares) and defect line (blue circles) regions (see Figures~S5--S7 for corresponding resonant FWM images). In the pristine sample, the FWM response exhibits a single optical transition at approximately 1.95~eV in both regions. This feature is attributed to a resonance near the A exciton energy, where the emitted $\omega_\text{FWM}$ is in resonance. This same characteristic feature persists in both the 0.4~at\% and 2.0~at\% doped samples but shows a systematic redshift to approximately 2.0~eV, strongly correlating with the $P_2$ peak energy previously observed in our PL measurements. Significantly, in the V-doped samples, excitations between 1.85 and 1.95~eV reveal an additional, distinct resonance near 1.88~eV in both monolayer and defect line regions. This feature is attributed to vanadium-induced states associated with the $P_1$ peak. It is worth to note that probing energies below 1.85~eV was limited by our optical setup. 

Overall, these results demonstrate the ability of resonance-enhanced FWM imaging to detect and characterize defect-related optical states in V-doped WS$_2$ monolayers. The unique resonance-enhanced signal offers direct insights into defect-induced excitonic transitions, highlighting FWM's high spatial sensitivity and capability to reveal local modifications in the nonlinear optical response. In our FWM images, the experimental spatial resolution is close to 500 nm by using a 0.95 NA objective (see methods), and the acquisition times ranges from a few to 30 seconds. Therefore, such characteristics reinforces FWM's imaging as potential tool for rapid and high spatial resolution for advanced defect characterization in doped 2D semiconductors.

\subsection*{Density functional theory calculations}
To elucidate the mechanisms underlying the observed optical phenomena, specifically the A exciton quenching, energy blueshift, and the unique FWM response, we performed Density Functional Theory (DFT) calculations (see Methods). Our analysis was conducted in two parts: first, examining the electronic structure induced by a singe vanadium substitution, and second, exploring the effects of increasing dopant concentration and disorder. 

\begin{figure}[!htbp]
    \centering  
    \includegraphics[width=1.0\textwidth]{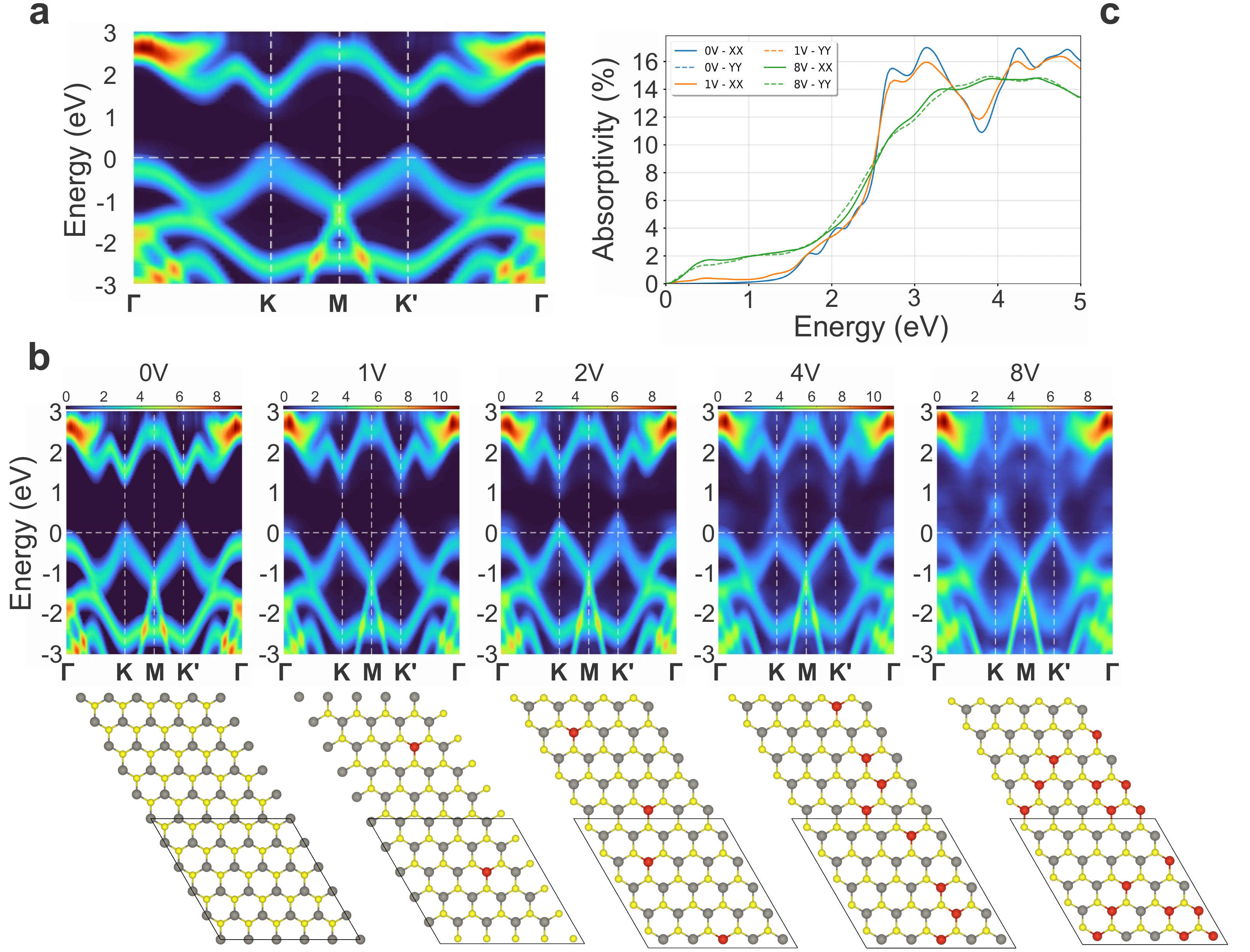}
    \caption{{\bf Density functional theory calculations of vanadium-doped WS$_2$ monolayers.}
    {\bf a.} Unfolded band structure of a $7\times7$ WS$_2$ supercell with a single V substitution (2\% concentration). Vanadium induces in-gap states and breaks valley symmetry. Color intensity indicates the primitive-cell spectral weight (Bloch character), while the small circles show the projection onto V states, with their size proportional to the spectral function weight. 
    {\bf b.} Top panel: Evolution of the unfolded electronic band structures, from semiconducting to metallic behavior, revealing the emergence of mid-gap defect states with increasing vanadium concentration. Bottom panel: Atomic structures of the $5\times5$ WS$_2$ supercells, showing the spatial distribution of vanadium dopants at varying concentrations (0, 1, 2, 4, and 8~V dopants).
    {\bf c.} Calculated 2D optical absorption spectra for pristine, one V, and eight V cases, showing the redshift in the onset of absorption and the emergence of anisotropy at high doping. The incident light is polarized along the zigzag (XX) and armchair (YY) directions, represented by solid and dashed lines, respectively.
    }
    \label{fig6}
\end{figure}

Our analysis begins with the substitution of a single tungsten (W) atom with a vanadium (V) atom in a $7\times7$ supercell, corresponding to a vanadium concentration of 2\%. As shown in Figure~\ref{fig6}a, the calculated unfolded band structure~\cite{PhysRevLett.104.216401} reveals the introduction of V-induced in-gap states, which is consistent with previous reports~\cite{sousa2024effects,review-apr,sousa2024giant}. We also observe a breaking of valley symmetry between the $\mathbf{K}$ and $\mathbf{K}^\prime$ points~\cite{sousa2024giant,review-apr}. The in-gap states are primarily associated with the $d_{z^2}$ orbitals of both V and W atoms (see Figures~S8--S13)~\cite{sousa2024effects,PhysRevB.88.085433}, and these impurity states are potential candidates for carriers traps that quench photoluminescence~\cite{rosa-WS2,review-apr}. Notably, the introduction of V also leads to a reduction of Bloch character in the calculated effective band structure, particularly near $\mathbf{K}$ valleys. This loss of Bloch state character would reduce quasiparticle lifetimes, thereby inhibiting radiative recombination pathways and providing a theoretical explanation for the observed photoluminescence quenching~\cite{PhysRevLett.104.236403,PhysRevB.85.085201,PhysRevB.90.115202,WANG2021108033}.

To examine the impact of higher doping concentrations and the inherent disorder in these systems, we employed the Special Quasirandom Structures (SQS) method~\cite{sqs,sqs2,icet} to model various vanadium concentrations ($x = 0.04~\text{to}~0.32$) within a $5\times5$ supercell. The formation energies, $E_f$, of the SQS structures was determined using the equation:

\begin{equation}
E_f=\frac{E_{\text{W}_{1-x}\text{V}_x \text{S}_2}-(1-x)E_{\text{WS}_2}-xE_{\text{VS}_2}}{\text{number of atoms}},
\label{eform}
\end{equation}

\noindent where $E_{\text{W}_{1-x}\text{V}_x \text{S}_2}$ is the total energy  of the disordered $\text{W}_{1-x}\text{V}_x \text{S}_2$ structure, $E_{\text{WS}_2}$ is the total energy of pristine WS$_2$, and $E_{\text{VS}_2}$ is the total energy of pristine VS$_2$, all for the same supercell. This quantity is equivalent to the enthalpy of mixing, $\Delta H_{m}$, of W$_{1-x}$V$_x$S$_2$ from the segregated phases at zero temperature. Our calculations show that the formation energies decrease from 358.8 to 216.3~meV per vanadium atom as $x$ increases from 0.04 to 0.32. This indicates a small reduction in energetic cost, or at least a consistent energetic cost, with increasing vanadium concentration.

Although the positive formation energies suggest that segregated pristine phases are more stable at zero temperature, a clearer perspective is provided by the mixing Helmholtz free energy~\cite{LIU2025162338}, $F_{m}=\Delta H_{m}-\Delta S_{m}T$. Here, $\Delta H_{m}=E_f$ is the mixing enthalpy, $\Delta S_{m}$ is the mixing entropy, and $T$ is the thermodynamic temperature. Since the mixing entropy increases with disorder, the Helmholtz free energy can become negative at elevated temperatures of the CVD growth process. This would lead to the spontaneous formation of disordered structures, providing a compelling theoretical explanation for the nanoscale doping inhomogeneity and preferential incorporation of vanadium observed in our experimental PL and SPM maps.

Figure~\ref{fig6}b shows the evolution of the electronic and optical properties of the system with increasing vanadium concentration. The supercell-projected band structures (see Figures~S8--S13), which incorporates contributions from both orbital and spin components, shows a well-defined Bloch character for the valence band maximum and conduction band minima of pristine WS$_2$. For low concentrations (e.g., 1V and 2V), a new band emerges below the pristine conduction band, characterized by states with limited Bloch character and visibility only at the $\mathbf{K}$ and $\mathbf{K}^\prime$ points. At higher concentrations (e.g., 4V atoms or more), a significant introduction of additional defect states occurs within the mid-gap region. Importantly, these mid-gap states becomes $\mathbf{K}-\mathbf{K}^\prime$ asymmetric, and the overall system begins to exhibit characteristics of a transition toward metallic behavior, with a non-zero spectral density across the entire pristine bandgap. This electronic transition directly explains the blueshift of the A exciton ($P_2$ peak) and its eventual quenching, as the strong Bloch character of the excitonic states is diminished.

The optical absorption spectra (Fig.~\ref{fig6}c) provide a key link to our FWM findings. The pristine system exhibits well-defined first and second optical transitions at 1.72~eV and 2.07~eV, with maximum absorption at 3.1~eV. Upon vanadium incorporation, the onset of absorption shifts to the infrared ($\approx 0.5$ eV), indicating that the V-induced mid-gap states are optically active. This optical activity, which is enhanced at higher doping concentrations, directly correlates with the strong, resonance-enhanced FWM signal at lower energies we observed in the defect-rich regions. In essence, the FWM technique selectively probes these new, optically active defect states that are otherwise masked in linear measurements by a quenched photoluminescence.

Our DFT calculations provide a comprehensive theoretical framework for the experimental observations. They confirm that vanadium substitution, together with associated disorder, introduces mid-gap states that break time reversal symmetry and reduce the Bloch character of excitonic states, thereby explaining the PL quenching and blueshift. Moreover, the calculations confirm that these defect states are optically active, providing a direct explanation for the strong, resonance enhanced nonlinear FWM signal observed exclusively in the defect rich regions of the doped WS$_2$ monolayers.

\section*{Conclusion} 
We have applied and demonstrated a resonance-enhanced FWM imaging methodology to characterize defect states in vanadium-doped WS$_2$ monolayers. Our multi-modal spectroscopic approach, which includes hyperspectral PL and Raman mapping, unveiled new excitonic features and spectral shifts indicative of nanoscale doping inhomogeneity. DFT calculations supported these findings by confirming that vanadium substitution introduces optically active mid-gap states, which reduce the Bloch character and explain the observed PL quenching and blueshift. A thermodynamic analysis also explains the spontaneous formation of the observed doping inhomogeneity. The FWM technique emerges as a powerful, rapid, and spatially sensitive alternative to linear spectroscopies, as it selectively probes these theoretically confirmed defect states. This work establishes FWM as a pivotal tool for advanced defect engineering in 2D semiconductors, holding immense promise for next-generation quantum photonics and optoelectronic devices.

\section*{Methods}
\paragraph*{Sample preparation.} 
The WS$_2$ monolayers, both pristine and vanadium-doped, were fabricated using a single-step chemical vapor deposition (CVD) process. The sample exhibited regular triangular shapes. By carefully controlling the concentrations of vanadium precursors during synthesis, flakes with average vanadium concentrations of 0.4~at\% and 2.0~at\% were obtained. 
The average atomic vanadium concentrations for these samples were previously characterized and reported in detail elsewhere~\cite{zhang-VWS2}.

\paragraph*{Raman and Photoluminescence measurements.} 
Raman and Photoluminescence (PL) spectra were acquired using a WITec Alpha 300 micro-Raman spectrometer. For Raman measurements, 457~nm excitations laser at $\sim$300~$\mu$W power and an 1800 lines/mm grating were employed with resolution better than 1 cm$^{-1}$. PL measurements utilized a 532~nm excitation laser at $\sim$150~$\mu$W power with a 600 lines/mm grating. Data collection was performed using an electron multiplying charge-coupled device (EMCCD). A Zeiss 100$\times$ objective (numerical aperture $=$ 0.9) was used for both laser focusing and signal collection. All measurements were conducted at room temperature at the LCPNano facilities at UFMG. Spectral analysis was performed using PortoFlow (FabNS) software. Raman spectra were fitted with Lorentzian functions for the $E^{\prime}$ and $A^{\prime}_{1}$ modes, and for the silicon substrate reference peak. The intensity of each spectrum was normalized based on the silicon peak's intensity and position. PL spectra were fitted using Gaussian function for the $P_{1}$ and $P_{2}$ excitonic peaks. 

\paragraph*{Fluorescence imaging.} 
Optical microscopy images (Figure~S14) were collected using an Nikon Eclipse upright microscope coupled with a Nikon fs-fi02 camera. A 60$\times$ objective lens with a numerical aperture (NA) of 0.85 was used for image acquisition.

\paragraph*{Nonlinear optical measurements.}
Four-wave mixing (FWM) experiments were conducted using an APE Berlin picoEmerald optical parametric oscillator (OPO) beam system. This system provides a tunable output ($\omega_1$) in the range of 750--950~nm (1.35--1.65~eV) and a fixed output ($\omega_2$) at 1064~nm (1.16~eV). The OPO operated with a 7~ps pulse width and an 80~MHz repetition rate. Backscattered nonlinear FWM signals were collected by the same objective used for excitation and directed to a photomultiplier tube (PMT) via dichroic mirror. FWM signals were generated at $\omega_\text{FMW} = 2\omega_1-\omega_2$, and by tuning $\omega_1$ in the 775--830~nm (1.49--1.60~eV) range, FWM emissions within 610--680~nm (1.82--2.03~eV) were detected, covering the WS$_2$ A exciton resonance. A 40$\times$ objective (numerical aperture $=$ 0.95) was used for beam focusing and signal collection. A specific set of optical filters was employed to select the desired FWM wavelength, as detailed in Table~S1 of the Supporting Information. Second Harmonic Generation (SHG) imaging, presented in Figure~S15 of the Supporting Information at 405~nm and 395~nm, was performed using the identical optical setup by collecting the SHG signal instead of FWM. For FWM intensity profile generation (Figure~\ref{fig5}), raw counts (intensity) from each FWM image at a specific wavelength were processed by subtracting the background response from the silicon substrate and normalizing by counts from a quartz reference standard. Consistency was ensured by maintaining the same laser wavelengths, incident power, and filter configurations throughout all FWM measurements. A schematic of the nonlinear optical experiment setup is shown in Figure~\ref{fig4}. The spatial resolution of the FWM images were determined by fitting the derivative of the intensity profile close to the edge of the WS$_{2}$ flake with a gaussian function and finding its full width at half maximum. 

\paragraph*{Theoretical calculation.}
All calculations were performed using the Vienna Ab initio Simulation Package (VASP) code, employing the projector-augmented wave (PAW) method~\cite{PhysRevB.50.17953,PhysRevB.59.1758,PhysRevB.54.11169,KRESSE199615} within the density functional theory with on-site Coulomb correlations (DFT$+$U) formalism~\cite{hohenberg1964inhomogeneous,PhysRev.140.A1133,PhysRevB.57.1505}. The exchange-correlation functional was described by the Perdew-Burke-Ernzerhof (PBE) parametrization~\cite{perdew1996generalized}. On-site Coulomb correlations were accounted for via $U_{eff}=3.0$~eV applied to the $d-$orbitals in vanadium atoms~\cite{Yun2020,Duong2019}. Spin-orbit coupling (SOC) was included in all calculations~\cite{PhysRevB.93.224425}. 
A plane-wave basis set with an energy cutoff of 400~eV was used, and the energy convergence criterion was set to $10^{-6}$~eV. Periodic boundary conditions were applied with a minimum separation of 13~\AA along the $z$-direction to eliminate interlayer interactions. The reciprocal space grid was constructed using $\mathbf{\Gamma}$-centered meshes: 8$\times$8$\times$1 for the 3$\times$3 supercell, 6$\times$6$\times$1 for the 5$\times$5 supercell, and a 3$\times$3$\times$1 for the 7$\times$7 supercell. 
To characterize the random alloy W$_{1-x}$V$_x$S$_2$, the Special Quasirandom Structures (SQS) method, implemented in the ICET software~\cite{sqs,sqs2,icet}, as employed. This method facilitates the accurate estimation of intrinsic properties of random alloys, while maintaining their physical characteristics, even at finite scales. A $5\times5$ supercell (25 tungsten sites) was constructed to model target vanadium concentrations ranging from 4\% to 32\%. For each concentration, the most representative structural configuration was selected for further calculations. The SQS structures were generated using cutoff radii of 13.5~\AA, 6.0~\AA, and 4.0~\AA to optimize the correlation functions of first, second, and third nearest neighbors, respectively. 
Band structure analyses and transition dipole moment calculations utilized the pyband and VASPKIT software packages~\cite{pyband,WANG2021108033}.

\paragraph*{Scanning probe microscopy measurements.} 
Atomic force microscopy (AFM) and lateral force microscopy (LFM) measurements were performed using a CSC37 No Al probe (MikroMash, with a nominal spring constant $k= 0.3$~N/m and resonant frequency $\omega_0= 20$~kHz). Electrostatic force microscopy (EFM) experiments were conducted with a HQ:XSC11 probe (MikroMash, $k= 7$~N/m and $\omega_0= 155$~kHz). All scanning probe microscopy experiments were carried out using a Nanoscope V MultiMode8 SPM (Bruker Instruments).

\section*{Supporting Information.} Supporting Information is available from the website or from the author.

\section*{Acknowledgments}
F.M., F.B.S., A.P.M.B., B.R.A.N. M.J.S.M., I.F.C., H.C., B.R.C., and L.M.M. acknowledge financial support from the Brazilian agencies CNPq and CAPES. They also acknowledge support from the Brazilian Institute for Science and Technology of Carbon Nanomaterials (INCT Nanocarbono) and the Rede Mineira de Pesquisa em Materiais Bidimensionais (FAPEMIG). 
M.J.S.M. and I.F.C. also acknowledge support from the Universidade Federal de Ouro Preto. The authors also acknowledge the support received from the National Laboratory for Scientific Computing (LNCC/MCTI, Brazil) at São Paulo (CENAPAD-SP) and the SDumont supercomputer at Rio de Janeiro.
B.R.C. also thanks the support provided by the Universidade Federal do Rio Grande do Norte.
M.L., D.Z., and M.T. thanks for financial support the AFSOR Grant No. FA9550-23-1-0447 and the National Science Foundation I/UCRC Phase II program at the Pennsylvania State University: Center for Atomically Thin Multifunctional Coatings (ATOMIC), Grant No. 2113864.
We are grateful to LCPNano for providing access to their facilities and equipment. We also thank FabNS for access to the PortoFlow software used for data analysis.

\paragraph*{Author contributions.} F.M., F.B.S., and L.M.M. conceived the study. F.M. and F.B.S. performed Raman, PL, FWM, and SHG measurements. A.P.M.B performed and B.R.A.N. analysed SPM experiments. M.L., and M.T. prepared the samples. F.M., F.B.S., B.R.C., and L.M.M. analyzed the optical data. F.M., F.B.S, B.R.C., and L.M.M. wrote the manuscript. B.R.C., and L.M.M. supervised the project. All authors reviewed and agreed on the final version of the manuscript.


\paragraph*{Competing Interests.} 
The authors declare no competing interests.


\bibliographystyle{nature}
\bibliography{references}

\end{document}